# Planar Cassegrain-type Schwarzschild Objective with Optical Metasurfaces

Xuan Liu[1†], Junhong Deng[1†], King Fai Li[1†], Yutao Tang[1], Mingke Jin[1], Jing Zhou[2], Xing Cheng[1], Wei Liu[3]* and Guixin Li[1,4]*

**Affiliations:**

[1]Department of Materials Science and Engineering, Southern University of Science and Technology, Shenzhen, China.

[2]Applied Optics Beijing Area Major Laboratory, Department of Physics, Beijing Normal University, Beijing, China.

[3]College for Advanced Interdisciplinary Studies, National University of Defense Technology, Changsha, China.

[4]Shenzhen Institute for Quantum Science and Engineering, Southern University of Science and Technology, Shenzhen, China.

*Correspondence to: wei.liu.pku@gmail.com and ligx@sustc.edu.cn

† These authors contributed equally to this work.

**Abstract:** Curved reflective mirrors play an indispensable role in widely-employed Cassegrain-type optics, which consequently are inevitably bulky, and thus neither integratable nor well extendable. Through substituting geometric phase based reflective metasurfaces for curved mirrors, here a planar Cassegrain-type Schwarzschild objective is obtained, with both the focusing and imaging functionalities experimentally verified. The planar reflective objective achieved is intrinsically free of residual light and thus imposes less stringent restrictions on illuminating light and nanofabrications compared to its refractive counterpart. Our planar Cassegrain-type designs can potentially reshape and simplify the core components of the microscope and telescope systems working at a broadband spectral range, and thus opens new avenues for designing compact imaging systems.

**One Sentence Summary:** A planar Schwarzschild objective is demonstrated, which can trigger an avalanche of reconstructions and refinements of reflective optical systems with metasurfaces.

Optical mirrors, either planar or curved ones, are probably the most ubiquitous optical elements in both everyday infrastructures and scientific apparatus (*1–5*). Similar to the irreplaceable roles played by lenses in refractive objectives, mirror-based optical elements serve as the most fundamental components in various reflective applications, which are essential for applications relying on microscopes, cameras, telescopes and other imaging systems (*2–5*). For more accessible and convenient visual observation and photography, the images are supposed to be on the opposite sides of the objectives with respect to the incoming electromagnetic waves. From this perspective, a single lens itself acts as the simplest refractive objective; while for reflective objectives, a conjugation of at least two mirrors are required, where the Cassegrain family of two-mirror configurations are most widely employed (*2–5*). Within all conventional Cassegrain-type Schwarzschild objectives, be them reflective, refractive, the efficient phase tuning and beam redirections rely on the continuous geometric curvatures of the consisting elements, which inevitably result in cumbersome and costly optical devices. This drops severe obstacles for many advanced applications requiring more compact optical apparatus, which are flexibly extendable for large-scale manufacturing and integration.

The situation has been totally changed with the explosively developing field of optical metasurfaces, which are composed of spatially variant meta-atoms on a planar interface and show unprecedented flexibility for optical wave manipulations (*6–8*). The planar configurations based on metasurfaces pervade redesigns for many optical elements, which are widely utilized for various applications, including holograms (*9–15*), polarization conversions (*16–21*), spin-orbital interactions (*22–29*), anomalous reflectors and optical retroreflectors (*7, 8, 30–32*), to name but a few. Among all the metasurface-based optical components, probably the most attractive and promising example is the planarized refractive objective lenses, which could consist of an individual metalens (*33–37*), metalens doublets (*38–40*), and other more sophisticated designs (*41–43*). The incorporation of flat metalens can significantly simplify many refractive optical elements, which then become more easily integrated and can play significant roles in ultra-compact optical devices with more advanced functionalities. Along with this flourishing trend, the unfortunate situation is that almost all attention has been focused on refractive objectives, with their reflective counterparts more or less sinking into complete oblivion. There do exist investigations into reflection-type metasurfaces (*7, 8, 44, 45*), nevertheless as mentioned above, a single reflecting element does not make an effective reflective objective that requires images and incoming light being on the opposite sides (*2–5*). Considering that for many optical devices reflective objectives are as essential as refractive ones, and within modern telescopes and lots of infrared systems the reflective designs are fully irreplaceable (*5, 46*), to make reflective optical components flat and compact are of paramount significance to revolutionize many frontiers of optical instrumentation.

Here we investigate the simplest and most widely employed two-mirror reflective configuration and realize a planar Cassegrain-type Schwarzschild objective, of which both the functionalities of focusing and imaging are experimentally demonstrated. The curved concave and convex primary and secondary mirrors in the conventional Schwarzschild objective are replaced by a pair of conjugated planar metasurfaces, which consist of judiciously oriented gold meta-atoms and perform converging and diverging optical functions, respectively. The wavefront shaping and beam routing within the metasurface objective are achieved by the efficient tuning of the geometric Pancharatnam-Berry (P-B) phase, which is inherent to a cascaded process of circular polarization transformations. Compared to the refractive metalenses based on geometric P-B phase metasurfaces, our reflective design is intrinsically free of residual light, and thus is less demanding for the circular polarizations of the incident wave. It is expected that our work here can trigger an avalanche of investigations into reflective optical components, which can potentially set off the trend to reconstruct and refine many optical devices including microscopes, spectrometers, telescopes, and so on that can be applied to observations ranging from microscopic to celestial scales over much broader electromagnetic spectral regimes.

**Designs of the planar Schwarzschild objective**

The conventional Cassegrain-type Schwarzschild objective is shown schematically in Fig. 1A. The objective consists of one concave primary mirror and one convex secondary mirror. A judicious combination of designed geometry and spatial parameters for the two mirrors leads to a reflective Schwarzschild objective, which exhibits both functionalities of focusing and imaging, and is superior to refractive lenses for specific applications (*46*). A planar version of the Cassegrain-type Schwarzschild objective is shown in Fig. 1B, where the curved mirrors are replaced by two metamirrors made of flat metasurfaces with correspondingly diverging and converging optical functions. The required phase distributions upon reflections by the metamirrors are associated with the geometric P-B phase (*34–37*), which is inherent to cascaded optical polarization conversions (*47, 48*). To be more specific, upon each reflection, the originally left-/right- circularly polarized (LCP/RCP) light is converted to RCP/LCP light. As a result, upon the two rounds of reflections by the metamirrors, at the opposite sides of the metasurface objective the optical wave is of the same circular polarizations (Fig. 1B). In both the conventional and planar designs there is an aperture through the primary mirror. Figures 1C and 1D are the two-dimensional cross-section views of Figs. 1A and 1B

The corresponding geometric parameters of the designed planar Schwarzschild objective are shown in Fig. 2A. The primary and secondary metamirrors (PM and SM) are off-set by $h$ along $z$ direction; the radius of SM is $R_1$, and the inner and outer radii of PM are $R_2$ and $R_3$, respectively. The focal length $f$ is defined as the distance between the focal spot and the PM; the numerical aperture (N. A.) of the objective equals to $n\sin\theta$, where $n$ is the index of the background medium (vacuum in this work), and $\theta = \arctan(R_3/f)$ is the central-axis deviating angle of the ray connecting the focal spot and the outmost rim of the PM. Based on the geometrical parameters, the required phase distributions upon PM and SM reflections are then given by:

$$\Phi_P(r_p) = -\int_0^{|r_p|} \frac{2\pi n}{\lambda} \cdot \left( \frac{r_2}{\sqrt{r_2^2 + f^2}} + \frac{r_2 - r_1}{\sqrt{(r_2 - r_1)^2 + h^2}} \right) dr_2, \qquad R_2 \leq |r_p| \leq R_3 \qquad (1)$$

$$\Phi_S(r_s) = \int_0^{|r_s|} \frac{2\pi n}{\lambda} \cdot \frac{r_2 - r_1}{\sqrt{(r_2 - r_1)^2 + h^2}} dr_1, \qquad 0 \leq |r_s| \leq R_1 \qquad (2)$$

where $\lambda$ is the wavelength of light in vacuum and $r_2 = r_1(R_3 - R_2)/R_1 + R_2$ (*49*).

To obtain metamirrors with both high cross-polarization conversion efficiency and the above-mentioned phase distributions, we employ a metasurface design with a ground gold plane, a $SiO_2$ dielectric spacer layer and a top layer of gold meta-atoms (length L, width W, and height H) with in-plane orientating angle φ, and the basic building block is shown in Fig. 2D. This metal-dielectric–metal configuration supports localized gap surface plasmon resonances (*50*), the absorption of which can be significantly reduced by optimizing the geometric parameter and then high optical efficiency can be obtained. In Fig. 2E we show the cross-polarization (LCP/RCP to RCP/LCP) conversion efficiency with normally incident circularly polarized light (FDTD solver, Lumerical Inc.) for a periodic configuration (refer to figure caption for more details of geometric parameters). It is clear that a cross-polarization reflectivity over 80% is obtained within a broad spectral range between 700 nm and 1000 nm. At the same time, each meta-atom with varying φ converts the incident circularly polarized light into that of opposite handedness. Intrinsic property of this process, an additional φ-dependent P-B phase shift is imparted into the light wave. As a result, the corresponding orientation angle distributions associated with the phase profiles specified in Eqs. 1 and 2 can be figured out directly (*49*). The fabricated PM and SM with the phase distributions shown in Figs. 2B and 2C are presented by direct photographs in Figs. 2F and 2G, and by scanning electron microscope images in Figs. 2H and 2I.

**Focusing properties of the planar Schwarzschild objective**

A planar Schwarzschild objective with the two fabricated metamirrors shown in Figs. 2F to 2I can focus an LCP wave to a focal spot with the same polarization, as is schematically demonstrated in Fig. 1B. The focusing properties are experimentally characterized (*49*). Firstly, a collimated LCP incident light wave has been focused by the planar metasurface objective; then the focal spot is magnified through a 10×objective lens and a tube lens with $f$=100 mm and finally is recorded with a CCD camera. The measured focal spot at $\lambda$=780 nm is shown in Fig. 3A, which is highly symmetric and can justify the excellent focusing property of the planar metasuface objective. A cross-section of the focal spot is shown in Fig. 3B, with a full-width at half-maximum (FWHM) of 3.16 μm that exceeds the estimated Abbe limit ($\lambda/(2\times\text{N. A.}) = 3.9$ μm). This seemingly super-resolution feature is induced by the obstruction of the secondary mirror (*51*), with an inevitable trade-off that a significant proportion of energy is distributed into the sidebands, which is quite similar to the case of optical super-oscillation (*52*). Figure 3C shows the measured field profiles in the axis plane with the PM located at $z = 0$. The measured focal position ($z = 2.2$ mm) is close to the designed value ($z = f = 1.99$ mm), and the deviation should originate from the effects of the substrate (that has not been considered in the theoretical analysis) and the fabrication imperfections. Furthermore, in Figs. 3D to 3E we also show the measured focusing properties at shorter wavelength of 633 nm. It is found that the measured focal length of the objective and FWHM of the focal point are 2.9 mm and 3.83 μm, respectively. This means that although the planar Schwarzschild objective is designed for a specific wavelength of $\lambda$=780 nm, it can also reasonably focus light at other wavelengths, though showing the spectral dispersion.

As the primary and secondary metamirrors have opposite dispersion characteristics with respect to diverging and converging functionalities, it is clear that a combination of them can reduce the chromatic dispersion of the whole objective. Here we characterize the chromatic dispersion by $\Delta f/\Delta \lambda$, where $\Delta f$ and $\Delta \lambda$ is the variation of focal length and wavelength, respectively. Numerical calculations show that the focal length of our designed metasurface Schwarzschild objective with N. A. = 0.1 would decrease with increasing wavelength from 600 nm to 900 nm, where approximately a linear dependence is exhibited (*49*). It is expected that the spectral dispersion can be further eliminated by optimizing the geometric parameters of the metasurfaces. Moreover, we have also measured the overall focusing efficiency, which is about 14% and 5% at wavelengths of 780 nm and 633 nm, respectively (*49*). The measured efficiencies is lower than that theoretically predicted one, however they can be improved by improving the processes of fabrication and optical alignment of the system.

**Imaging properties of the planar Schwarzschild objective**

To characterize the imaging performances, another metasurface Schwarzschild objective with $f = 10$ mm and a smaller numerical aperture N.A. = 0.05 (with broader field of view) has been designed and fabricated. The focusing properties of the new objective are presented in Figs. 3G to 3L, where it is shown that the metasurface Schwarzschild objective has the FWHM of the focal spot of 5.83 μm and 5.98 μm, and focal lengths of 9 mm and 11.3 mm, at wavelengths of 780 nm and 633 nm, respectively.

To perform the imaging experiment, we firstly deposit an 80 nm thick gold film on glass substrate, then mill three slits into the gold film utilizing focus-ion beam technique. As shown in Figs. 4A and 4D, the center to center distance of the slits are 10 μm and 8 μm with a filling factor of 0.5. Then, the object is placed at the focal plane of the metasurface Schwarzschild objective and illuminated by the focused laser beams with N. A. = 0.25 (*49*). Figures 4B, 4E and 4C, 4F show the images captured on the CCD camera at wavelengths of 780 nm and 633 nm, respectively. Due to the limited area size of focal spot of laser, only two slits are observable. For illumination laser at wavelengths of 780 nm and 633 nm, the images of the slits are magnified approximately by 20 and 10 times in the imaging system consisting of the planar Schwarzschild objective and a tube lens ($f$ = 100 mm). The images obtained prove that our designed metasurface Schwarzschild objective can image an object with a resolution up to 8 μm, which is more or less consistent with the measured FWHM of the focal spot in the focusing experiment (Figs. 3H and 3K).

Besides being comparable to its refractive counterpart of lenses or metalenses in terms of focusing and imaging performances, the reflective planar metasurface objective we have obtained shows its superiority to be intrinsically free of residual light. Upon reflection on each metamirror of the objective, those parts of waves that maintain the circular polarization or show the opposite handiness of circular polarization are directly reflected out of the objective and thus no residual light remains to degrade the focusing or imaging performances (*49*). In sharp contrast, those refractive metalenses that reply on geometric Pancharatnam-Berry phase cannot directly eliminate the residual parts and thus extra polarization analyzers are required, which inevitably reduces the compactness of the whole optical system (*34–37*).

**Conclusions and Outlook**

We have successfully incorporated flat optical elements made of plasmonic metasurfaces into classical Cassegrain-type reflective systems and demonstrate a planar Schwarzschild objective that can be applied for focusing and imaging applications. Though the reflective objective obtained stills suffers from chromatic aberrations and low efficiencies, those problems can be solved through further structural and geometric optimizations. We believe our work opens the door to metasurface-based revolutionary refinements and miniaturizations for not only refractive but also reflective optical systems. The fusion of refractive optical elements and planar meta-optics paves new avenues for manufacturing and reconstructing more efficient microscopes, telescopes and other optical devices over much broader spectral regimes, and can potentially accelerate the developments of many related fields including artificial intelligence photonics, virtual reality optics, and other wearable and portable optical sensors and detectors.

**Acknowledgments:**

**Funding:** Supported by the National Natural Science Foundation of China (Grant No. 11774145), Applied Science and Technology Project of Guangdong Science and Technology Department (2017B090918001) and Natural Science Foundation of Shenzhen Innovation Committee (JCYJ20170412153113701). **Author contributions:** All authors wrote the paper and participated in discussion. X. L., J. D. and K. L. contribute equally. **Competing interests:** Authors declare no competing interests. **Data and materials availability:** All data is available in the main text or the supplementary materials.

**Supplementary Materials:**

Materials and Methods

Figures S1-S7

References

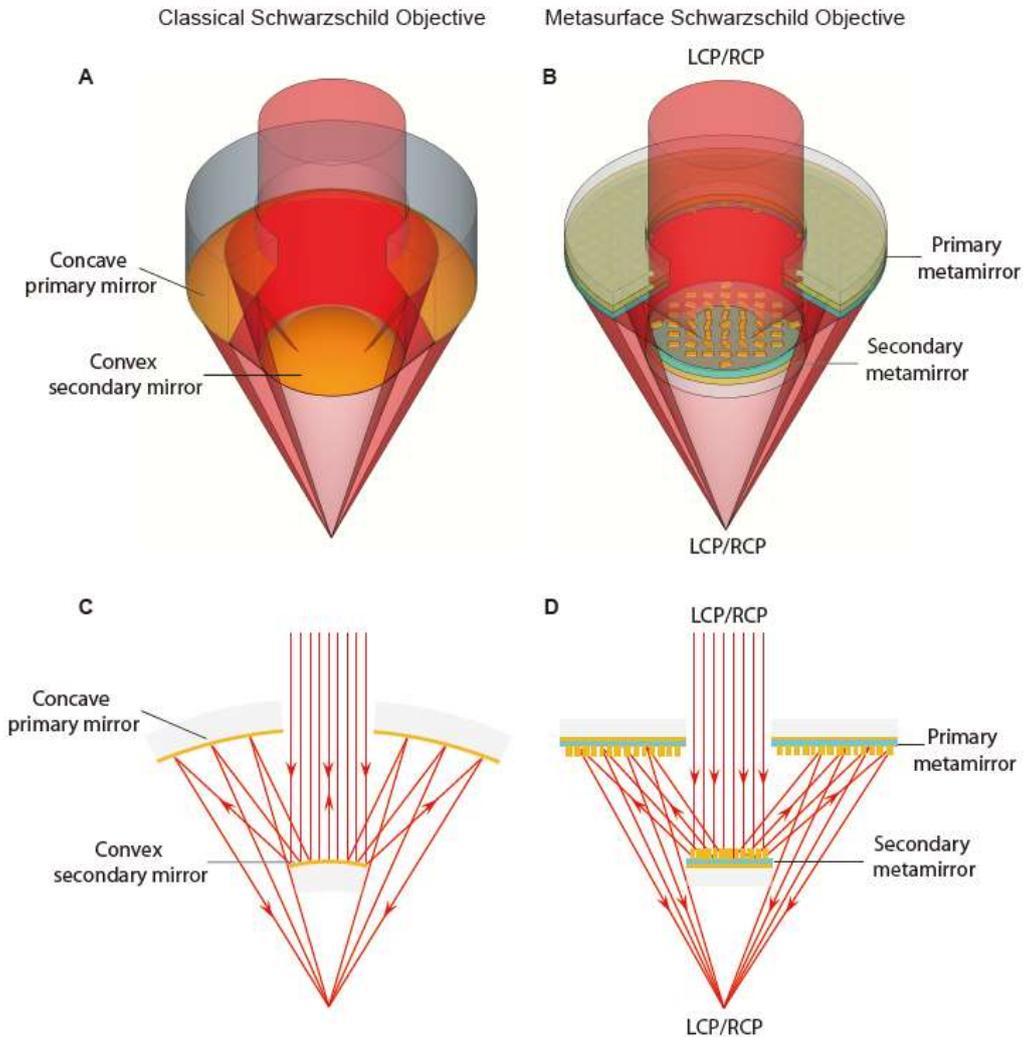

**Fig. 1. Schematic illustration of curved and planar Cassegrain-type Schwarzschild Objectives.** (**A**) The conventional objective, consisting of a concave primary mirror and a convex secondary mirror, focuses collimated incident light after being reflected twice. (**B**) The miniaturized planar counterpart of the conventional Schwarzschild objective, where the conventional curved mirrors are replaced by planar metamirrors with pre-designed gradient phase distributions. The required phase profiles are realized based on geometric Pancharatnam-Berry phase, which enables the diverging and converging functionalities for the secondary and primary metamirrors, respectively. Upon the two rounds of reflections upon metamirrors, the originally circularly polarized incident light is focused with the same circular polarization state. (**C**) and (**D**) are the two-dimensional cross-section views of objectives in (A) and (B), respectively. The red lines represent the trajectories of light rays.

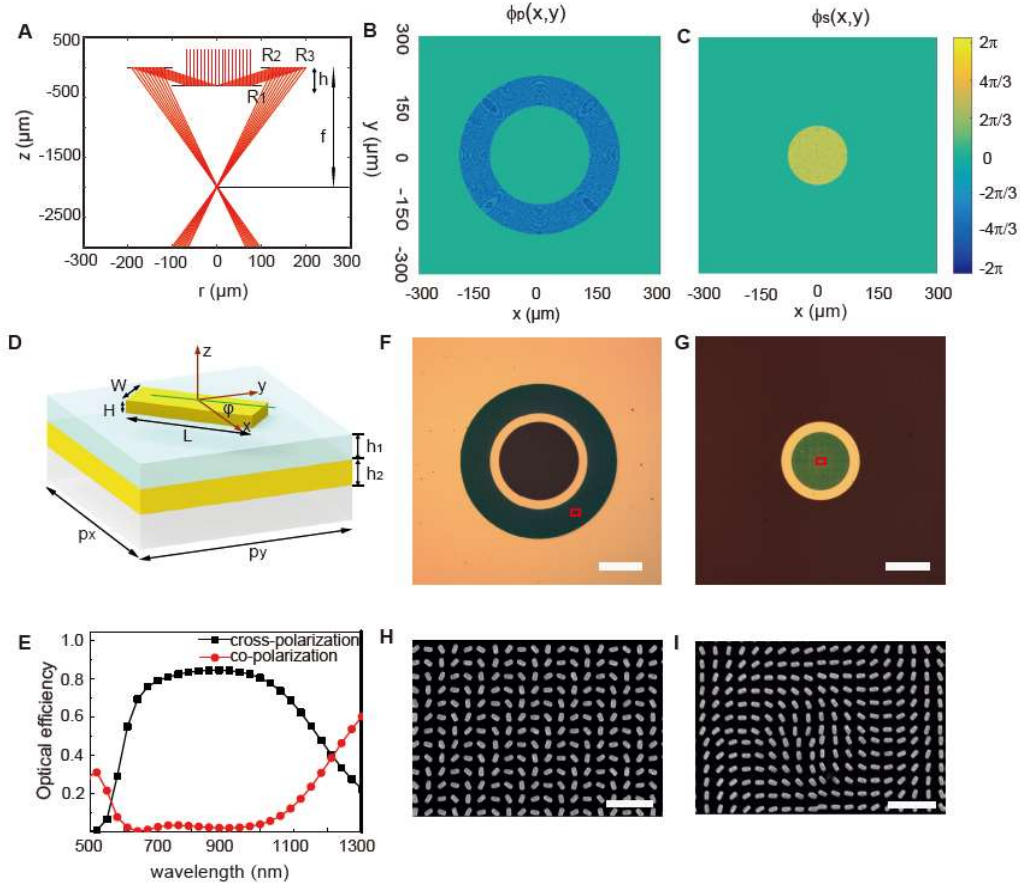

**Fig. 2. Design and fabrication of the planar Cassegrain-type Schwarzschild Objective.** (**A**) Ray-tracing diagram of the planar metasurface objective. (**B**) and (**C**) The phase profiles required for the primary and second metamirrors (PM and SM) with geometric parameters: $R_1$ = 75 μm, $R_2$ = 124 μm, $R_3$ = 200 μm, h = 300 μm, f = 1.99 mm, NA = 0.1 and $\lambda$ =780 nm. (**D**) Building block for the metamirrors: glass substrate is coved by $h_2$ =100 nm thick gold film and $h_1$ = 100 nm thick SiO$_2$ spacer layer; the gold meta-atom on top of the SiO$_2$ film has length L = 200 nm, width W = 90 nm and height H = 30 nm, φ represents the orientation angle of the meta-atom in the x–y plane. The periods along the x and y directions are P$_x$ = 300 nm and P$_y$ = 300 nm, respectively. (**E**) The numerically calculated polarization conversion efficiency for circularly polarized light upon reflections by a unit cell of the metasurface. (**F** and **G**) The optical images of the fabricated PM and SM (scale bar: 100 μm), respectively. (**H** and **I**) SEM images (scale bar:1 μm) of the PM and SM over the regions marked by red boxes in (F) and (G), respectively.

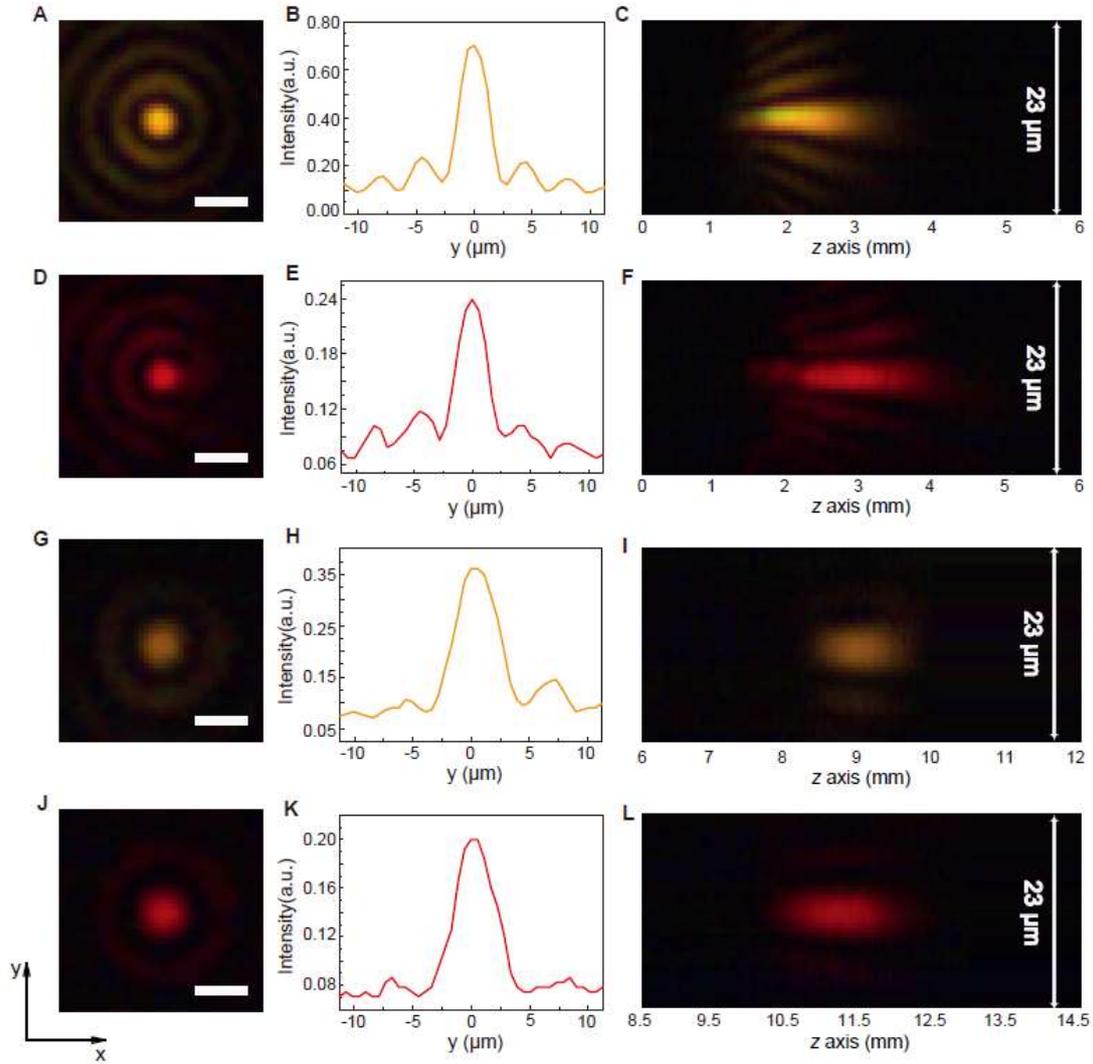

**Fig. 3. Focusing properties of the planar Cassegrain-type Schwarzschild Objective.** N. A. = 0.1 in (A) - (F); N. A. = 0.05 in (G) - (L). (**A**, **D**, **G** and **J**) The measured field intensity profiles at the focal plane (scale bar: 5μm). (**B**, **E**, **H** and **K**) The corresponding cross sections of the focal spots along *y* direction, with FWHM of 3.16 μm, 3.83 μm, 5.83 μm and 5.98 μm, respectively. (**C**, **F**, **I** and **L**) The corresponding intensity profiles along the propagating axial plane. In (A) - (C) and (G) - (I) the incident wavelength of light is 780 nm, while in (D) - (F) and (J) - (L) it is 633 nm.

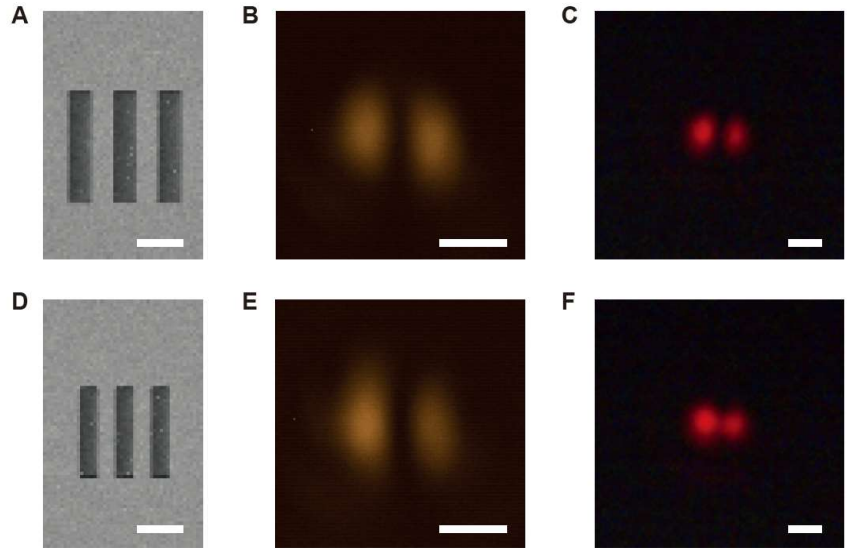

**Fig. 4. Imaging with the planar Cassegrain-type Schwarzschild Objective (N. A. = 0.05). (A and D)** The scanning electron microscopy images of the objects (scale bar: 10 μm), which are milled in an 80 nm thick gold film on glass substrate. The center to center distance of the slit is 10 μm (A) and 8 μm (D), respectively. The dark areas are transparent and have width half that of their separations, i. e. 5 μm (A) and 4 μm (D). **(B)** -**(C)** and **(E)**--**(F)** are images of the gold slits (scale bar: 10 μm) recorded by a homemade imaging system consisting of the metasurface objective, a tube a lens and the CCD camera. The wavelength of illumination laser for the two objects are $\lambda$=780 nm (B and E) and 633 nm (C and F), respectively. The magnification ratios of the imaging system are ~ 20 and ~ 10 times for illumination wavelengths at 780 nm and 633 nm, respectively.